\documentclass[reprint,aip,jap,amsmath,amssymb,nobalancelastpage,citeautoscript,floatfix,longbibliography, raggedbottom]{revtex4-2}
\usepackage[usenames,dvipsnames]{color}
\usepackage{graphicx,xfrac,microtype}
\usepackage[bookmarks=false,colorlinks]{hyperref}
\hypersetup{hypertexnames=false,linkcolor=RubineRed,citecolor=Plum,filecolor=Mulberry,urlcolor=RoyalBlue}
\usepackage{multirow}

\newcommand{\degC}[0]{$^{\circ}$C}

\newcommand{\Autoref}[1]{%
  \begingroup%
  \def\sectionautorefname{Appendix}%
  \autoref{#1}%
  \endgroup%
}

\begin{document}

\title{Benchmarking Structural Evolution Methods for Training of Machine Learned Interatomic Potentials}

\author{Michael J.\ Waters}
\affiliation{Department of Materials Science and Engineering, Northwestern University, Evanston, Illinois  60208, USA\textcolor{white},{$ $}}

\author{James M.\ Rondinelli}
\email{jrondinelli@northwestern.edu}
\affiliation{Department of Materials Science and Engineering, Northwestern University, Evanston, Illinois  60208, USA\textcolor{white},{$ $}}

\begin{abstract}
When creating training data for machine-learned interatomic potentials (MLIPs), it is common to create initial structures and evolve them using molecular dynamics to sample a larger configuration space.
We benchmark two other modalities of evolving structures, contour exploration and dimer-method searches against molecular dynamics for their ability to produce diverse and robust training density functional theory data sets for MLIPs. 
%
We also discuss the generation of initial structures which are either from known structures or from random structures in detail to further formalize the structure-sourcing processes in the future.
The polymorph-rich zirconium-oxygen composition space is used as a rigorous benchmark system for comparing the performance of MLIPs trained on structures generated from these structural evolution methods.
Using Behler-Parrinello neural networks as our machine-learned interatomic potential models, we find that contour exploration and the dimer-method searches are generally superior to molecular dynamics in terms of spatial descriptor diversity and statistical accuracy. 
\end{abstract}

\maketitle

\section{Introduction}

Simulations based on density functional theory (DFT) and semi-empirical interatomic potentials are often used to model the potential energies of atomic systems to interrogate static and dynamical phase evolution at the atomic scale, macroscopic physical properties, and finite-temperature effects by employing thermodynamic ensembles with molecular dynamics (MD). 
DFT has the advantage of providing quantum accuracy from first principles, but it is constrained to small system sizes due to its $O(n^3)$ scaling\cite{Artrith2016}. 
Interatomic potentials conversely have limited accuracy due to dependency upon the functional form and empirical parameterization used to describe the interactions between atoms, but they scale linearly. 
In recent years, machine learned interatomic potentials (MLIP) have demonstrated their ability to combine the best of both techniques\cite{Miwa2018, Behler2016}. 
Total energies are learned as a sum of atomic contributions that depend upon their local chemical environment from a training set of \textit{ab initio} calculations\cite{Behler2007, Behler2014}. 
There have been a number of studies looking at both improved descriptors of atomic environments\cite{Behler2016, Kocer2019, Goryaeva2019, Batra2019} and machine learning models\cite{Behler2016, Schmitz2019MachineMethods}. 
However, existing state of the art methods still rely upon a user supplying training samples that sufficiently sample the potential energy surface (PES) of interest. 
The \textit{de facto} standard is to use \textit{ab initio} MD to evolve an initial structure and sample the local configuration space. 
A key advantage of sequential evolution reduces computational cost and improved stability by preconditioning the electronic solver for subsequent steps.
%
\textit{Ab initio} MD typically does not require additional specific information about the system to successfully iterate through a specified number of trajectories; it is essentially structure-agnostic calculator and the same simulation parameters can be employed with any initial structure.
Furthermore, MD can stochastically samples structures from the canonical thermodynamic ensemble.
At first glance, sampling a physically meaningful distribution is appealing, however this offers no guarantee of efficiency in training a MLIP model. 
Additionally, rare events are inherently under-sampled. But rare events are the dominant feature of important physical processes like diffusion and chemical reactions.
These processes are controlled by PES saddle points, which are dynamic unstable points, and hence are difficult sample with MD. 
These issues still afflict active learning methods and initial data sets generation for adversarial learning methods since any new structures added to the training set would be obtained from the distribution sampled by MD. 
%
Hence, we are motivated to explore alternative dynamics for their passive improvements in data set generalizibilty before we consider more sophisticated learning techniques.

In this work, we examine the contour exploration (CE) technique\cite{contour_paper}, the dimer-method search\cite{dimer_vtst}, in addition to MD to generate structures for subsequent training of MLIPs. 
The CE technique evolves a system at a fixed target potential energy using curvature based extrapolation in the Frenet frame to accurately follow the potential energy contour. 
%
We hypothesize that CE should improve upon MD in two ways. 
The first is by removing the ambiguities of temperature-dependent and time-step dependent sampling in favor of directly comparable energy targeted sampling. 
The second area of improvement is that CE can reliably take much larger steps between iterations, which should reduce sampling redundancy.
Owing to its adaptive step size scheme, CE can be tuned to preferentially sample regions near PES saddle points.
This sampling, however, occurs at the expense of smaller steps and is possible only when the potential energy target is near the saddle energy.\cite{contour_paper}  
The dimer-method enables better and direct sampling of saddle points.
It is a minimum eigenmode-following method which optimizes to saddle points on the PES in an analogous way that optimization methods find minima.\cite{dimer_vtst} 
We mention that the dimer method (DM) is saddle point searching technique, so the trajectories/simulations are sometimes referred to as dimer searches.

Although we focus on benchmarking sampling dynamics choices, one cannot ignore the importance of the \emph{initial} structure generation for creating robust training data.
Generally, there are two categories of training structures: (1) polymorph structures informed by or derived from known phases and crystal structure types, and (2) random or glass-like structures, which are less or not at all informed by \textit{a priori} knowledge. 
Type (1) structures are employed, because we want MLIPs to accurately model realistic materials that are close to ideal structures of interest.  
In general, the list of known phases tends to contain the ground state structures which informs models of the minimum energy bounds. 
For example, it would be difficult to train a MLIP for iron without including face-centered cubic (FCC) and body-centered cubic (BCC)  structures in the training set.
Additional training structures, which we refer to as polymorph-derived structures, are also needed to sample interactions involving atomic substitutions, omissions, additions, and distortions, because real materials exhibit nonidealities arising from defects, phase inhomogeneity, and stresses.
Type (2) structures mimic amorphous solids, liquids, and gases to sample greater and more diverse local configuration spaces. 
This adds robustness to MLIPs when dealing with metastable structures, but tends to have poor representation of ground state crystal structures.  
A mixture of both structure types should yield the most accurate and robust training set. 
There are complexities in developing robust structure generation schemes for both Type (1) and (2) structures, which we discuss in detail, and our code implementation for automatically creating them aims to resolve.

We perform our analyses of the evolutionary methods for sampling structural spaces to generate data for machine-learned learned interatomic potentials in the zirconium-oxygen system, because of its industrial importance and the many stable phases accessible in the binary system provide increase complexity. 
We find that both CE and DM dynamics produce more diverse training data than MD. 
The dimer method produced the highest diversity of data, while in cross-evaluation the contour-exploration-trained MLIP was the most robust.
Since both methods have fewer tunable parameters than MD and more efficiently sample atomic configuration spaces, either one or a mixture of both, will be a more computationally efficient means of creating training data for MLIPs.

\section{Material and Methods}

\subsection{Use Case: Zr-O System}
Zirconium alloys are widely used in nuclear fuel cladding\cite{Motta2015}, dentistry\cite{Daou2014}, fuel cells\cite{Shim2007}, and gas sensors\cite{Maskell2000ProgressSensors}. 
Zirconium metal adopts a HCP structure ($\alpha$-Zr) up to 863 \degC then changes to BCC ($\beta$-Zr). 
Zirconia occurs in three polymorphs at atmospheric pressure: monoclinic (m-ZrO$_2$)\cite{Aldebert1985} at ambient, tetragonal (t-ZrO$_2$)\cite{Teufer1962TheZrO2} above 1170\,\degC, and cubic (c-ZrO$_2$) above 2370\,\degC \cite{Howard1988StructuresDiffraction, ZrO2_transitions}. 
At higher pressures, other orthorhombic polymorphs (o-ZrO$_2$)\cite{Kisi1989CrystalZirconia, Jain2013} have been reported as well.
With such a rich phase-space in ZrO$_2$ alone, several other authors have already used it as a benchmark for their MLIP methods.\cite{KresseZirconia, Wang2018}
Finally, a number of stable suboxide compounds are known, including ZrO, Zr$_2$O, and Zr$_3$O.\cite{avdv_ZrO}

\subsection{Energy and Force Calculations}
DFT calculations were performed using the Vienna \textit{ab initio} simulation package\cite{kresse1996}  (VASP,version 5.4.4) with the Perdew-Burke-Ernzerhof (PBE) generalised gradient approximation.\cite{perdew1996}
The Zr\_sv (04Jan2005) and O (08Apr2002) projector augmented wave (PAW) potentials\cite{blochl1994} were used to describe the core electrons. The plane wave cutoff energy was set to 500 eV. 
The $k$-point grids were automatically computed using  Monkhorst-Pack grids which maintained a minimum 10,000 $k$-points$\cdot$\AA$^{3}$. 
A conservative 8\% maximum electronic density mixing was used between Davidson-block iterations to ensure convergence of the self-consistency cycle across a wide variety of structures. 
VASP input files were automatically generated using the Atomic Simulation Environment's (ASE)\cite{ase-paper} interface to to VASP.

\subsection{Training Machine-Learned Interatomic Potentials}

We employ a Behler-Parrinello type neural network (NN) where the total energy of the system is modeled as a sum of individual atomic energy contributions. 
Each contribution is predicted using an element-specific NN, where the inputs to the NN are atom-centered spatial descriptors that encode the local chemical environment. 
For the spatial descriptors, we use Behler-Parrinello style Gaussian descriptors of types $G^{(2)}$ (2 atom), $G^{(4)}$ (3 atom), and $G^{(5)}$ (3 atom) as described in \Autoref{sec:descriptors} and from a fork of the Atomistic Machine-learning Package (AMP, version 0.7.0-beta)\cite{Khorshidi2016}  with added bug fixes documented at \url{https://bitbucket.org/koysean/ramp/src/master/}.

\begin{table}
\centering
\caption{The Gaussian-type spatial descriptor parameters used. $\eta$, $\zeta$, and $\gamma$ are unitless. The unique permutations for two elements results in 16 $G^{(2)}$ pairs, 96 $G^{(4)}$ triplets, and 24 $G^{(5)}$ triplets.}
\begin{ruledtabular}
\begin{tabular}{lllll}
$R_c$  & 7.0 \AA &  &   &   \\
\hline
$G^{(2)}$ $\eta$  & 0.781 & 0.1.479 & 2.800  &  5.300 \\
          & 10.0324 & 18.991 & 35.951 & 68.056 \\
\hline 
$G^{(4)}$ $\eta$   & 0.781 & 1.665  & 3.548  &  7.562   \\
$G^{(4)}$ $\zeta$  & 1.0   & 2.154  & 4.642  & 10.000\\
$G^{(4)}$ $\gamma$ & -1.0   & 1.0    &       &       \\
\hline
$G^{(5)}$ $\eta$   &  0.781 & 7.562  &        &       \\
$G^{(5)}$ $\zeta$  & 1.0       & 10.0  &       &        \\
$G^{(5)}$ $\gamma$ & -1.0      & 1.0   &       &       \\
\end{tabular}
\end{ruledtabular}
\label{tab:descriptor_set}
\end{table}

Equivalent sets of spatial descriptors
were used for all atoms, and the parameters are shown in \autoref{tab:descriptor_set}. 
With the unique permutations of the descriptor parameters and  $\eta$, $\zeta$, and $\gamma$, chemical elements, there were 24 $G^{(5)}$ triplets, 96 $G^{(4)}$ triplets and 16 $G^{(2)}$ pairs for a total of 136 descriptors per atom. 
A cosine cutoff function was used to ensure the descriptors and their derivative smoothly vanish outside of the cut-off radius ($R_c$). 
The neural network used for each element consisted of a single hidden layer of eight nodes all using hyperbolic tangent as the activation function. 
All MLIPs were trained using 50 trajectories starting from derived structures and 50 trajectories starting from random structures to give the most general training set.
Structure creation is fully discussed in Section D. 
The NN were trained using a loss function which included RMSE of energy, the RMSE of forces weighted to 10\%, and regularization/overfit penalty which was optimized using BFGS. 
For the first 200 iterations, the overfit penalty was 0.05 before being reduced to 0.001 for 1500 more iterations. 
The initially high overfit penalty prevented the NN from becoming trapped in local minima of the loss function but was reduced to avoid oversoftening of the atomic interactions.

\subsection{Structure Creation}

\subsubsection{Implementation}
Structure generation code used in this work is 
built upon the widely used Atomic Simulation Environment (ASE) package\cite{ase-paper} to ensure interoperability with many DFT and molecular mechanics software packages and wide compatibility with atomic data file formats. 
Like ASE, our code is written in Python due to its clear syntax, large scientific user base, and many free open-source libraries. 
It is available at \url{https://github.com/MTD-group/amlt/}. 
To use it, a user specifies known input structures and parameters in a Python script before running it. 
The output includes structures and input files needed to run DFT calculations for the training set.
%
Structures are represented as ASE \texttt{Atoms} objects, which ensures compatibility with a large number of different file formats.

\subsubsection{Limiting which Energies and Forces to Reproduce}
\begin{figure}
\centering
\includegraphics[width=0.45\textwidth]{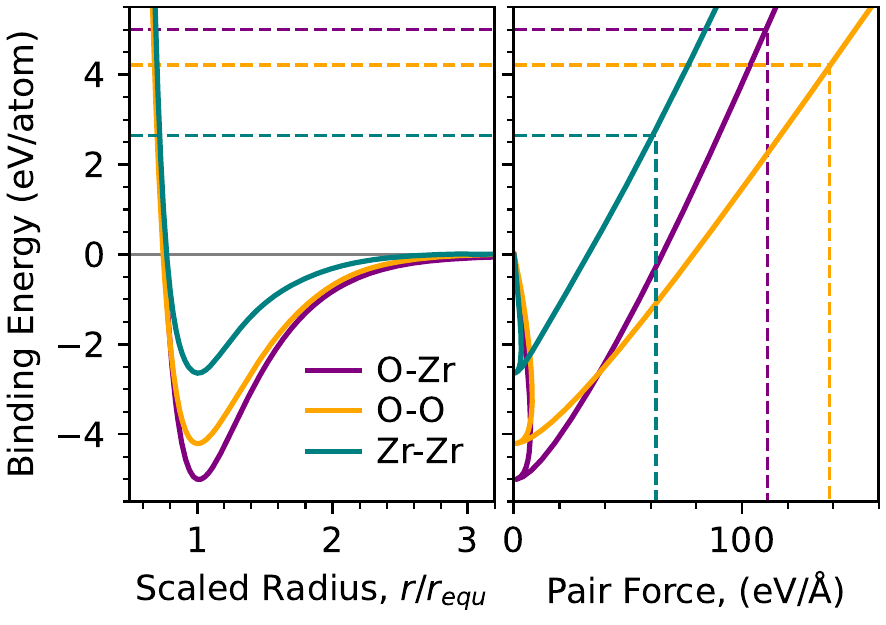}
\caption{The interatomic potential energy is plotted vs. radial separation scaled to the equilibrium distance for pairs of atoms isolated in a vacuum. By tracing the maximum interatomic potential energies of interest into the beak-like plot of energy vs. force plot, a maximum force of interest can be found. In this work, twice the binding energy above the potential well minimum is used.}
\label{fig:beak_plot}
\end{figure}

Towards the aim of reliably automating robust MLIP generation, it is clear not all structures are necessary or even helpful when parameterizing atomic interactions.
The inclusion of high-energy structures corresponding to high-atomic overlap helps establish the strong repulsive component of the interatomic potential; however, structures with excessively high energies and forces are drastically over-represented in loss-functions when these regions only need to be approximately correct for most applications outside of high-pressure research. 
Thus the question becomes how much of the repulsive part of the IP is needed? Or conversely, how do we filter/limit structures in a way that also easily conveys the upper limits of energy and forces? 
We propose a simple limit based on the binding energies of the different atomic species in the composition space. 
Using the energy versus distance curves of the elemental pairs (\autoref{fig:beak_plot}), and tracking the repulsive part up to a multiple (2X here) of the binding binding energy above the minimum, we can find an upper energy limit of interest. 
By plotting energy versus force, we also obtain a maximum force of interest. 
The maximum forces are useful for filtering out structures with many atoms that may contain a pair of highly overlapped, high-energy atoms.
In this case, average energy per atom is not sufficient for filtering since having many other atoms will dilute the high-energy contributions of the pair.
Thus, we propose using a multiple of the binding energy converted to upper limits on atomic forces through the bird beak-like energy--force plots.
This approach allows us to standardize the upper limits on energy and forces in an easy to compute and interpretable way. 
The computational cost for two-atom pairs is minimal and radial spacing of the atoms can be coarse since the near-linear behavior in the beak plot in \autoref{fig:beak_plot} permits accurate interpolation. 
In our case, we use 111 eV/\AA, because we are predominantly interested in the Zr-Zr and Zr-O interactions in the metal and oxides.

\subsubsection{Polymorph-Derived Structures}
\begin{figure}
\centering
\includegraphics[width=0.45\textwidth]{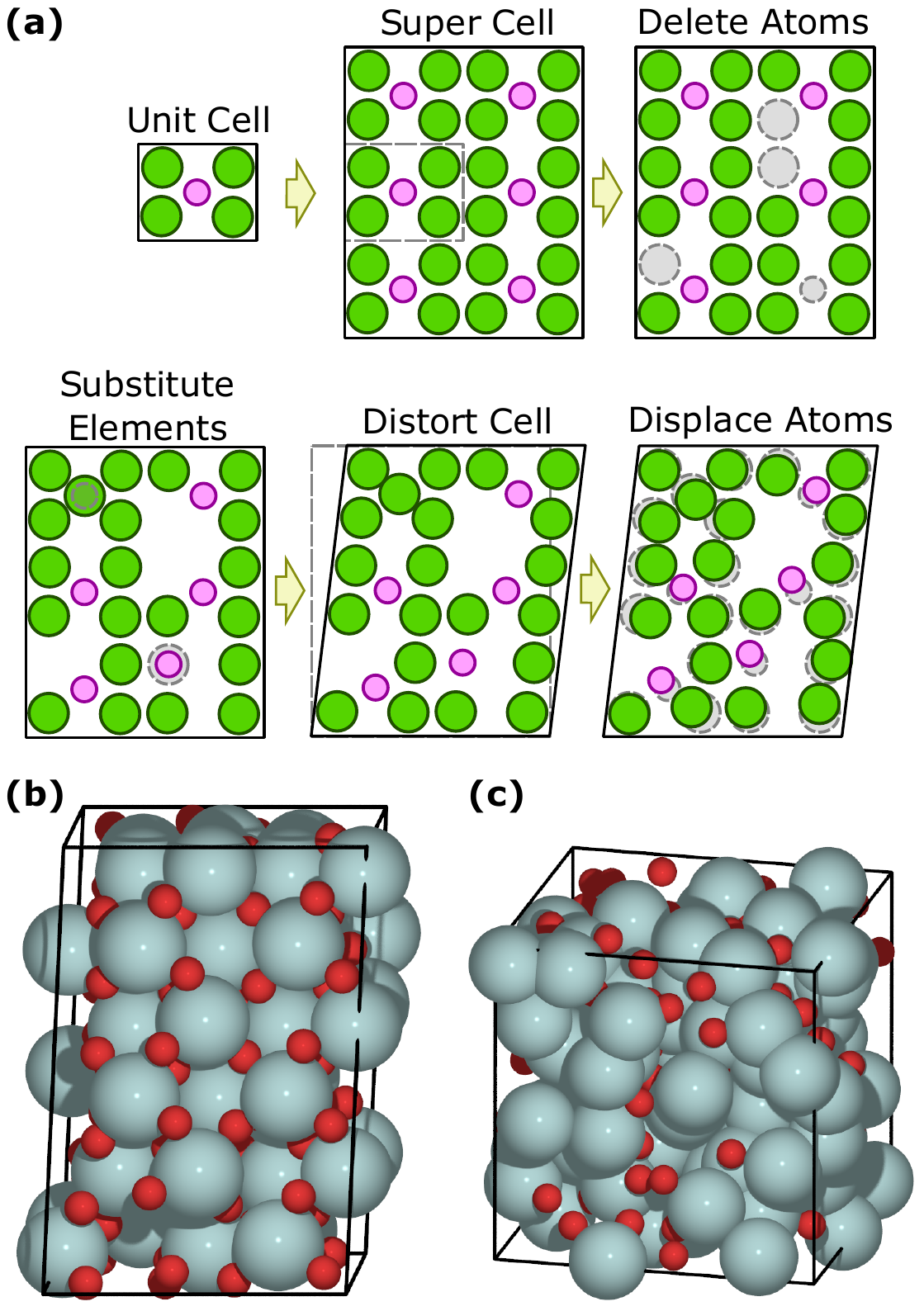}
\caption{(a) The process of creating a polymoprhous distorted super cell with random atomic displacements and deletions from the supplied unit cell of a known phase. (b) An example polymorph-derived structure derived from c-ZrO$_2$. (c) An example of a random structure in the Zr-O system. Small (red) and large (gray) spheres correspond to O and Zr, respectively.}
\label{fig:structure_examples}
\end{figure}

Phenomena studied with interatomic potentials typically involve perturbations and/or permutations of known crystal polymorphs such as lattice defects, impurities, phonons, displacive transformations, bulk strain, etc. 
%
To account for these effects not well described by random structures, we create polymorph-derived structures from a set of known phases starting from their primitive cells as illustrated in \autoref{fig:structure_examples}.
Our procedure starts with applying a random supercell operation to the primitive cell such that any of the lattice directions may be replicated $N$ times until it reaches a specified size limit. 
We choose the size limit such that it reflect the interaction cut-off distance. 
Next, atoms are randomly deleted from the supercell to introduce vacancies followed by a random substitution of chemical species which reflects alloying or other ordering defects. 
Next, the cell vectors are distorted in two steps. The first step works to apply a strain to the structure by finding a set of strained lattice vectors that change volume only within a small range. The strained lattice vectors are found by repeatedly applying proportional random distortions of the cell vectors until a new set of acceptable cell vectors is found. This procedure helps ensure that shearing of the cell does not cause drastic overlapping of atoms. The second cell operation is a bias-able dilation to reflect isotropic pressure effects. 
The final step is random atom displacements. 
In total, there are eight mutation parameters where seven are scale-free.
For our polymorph-derived structures, we include the primitive cells of $\alpha/\beta$-Zr (HCP/BCC), monoclinic, tetragonal, and cubic ZrO$_2$. 
We used a 5\% atom deletion chance to create vacancies and a 5\% substitution chance since the Zr-O system exhibits limited if any substitutional solubility. 
The random strain and dilation parts of the cell distortion were limited to a range of 10\% of the lattice vector magnitudes. 
The dilation part of cell distortion was biased 75/25\,\% tension/compression. Random individual atomic displacements were limited to 0.2 \AA. 
%
%

\subsubsection{Random Structures}
\begin{figure}
\centering
\includegraphics[width=0.45\textwidth]{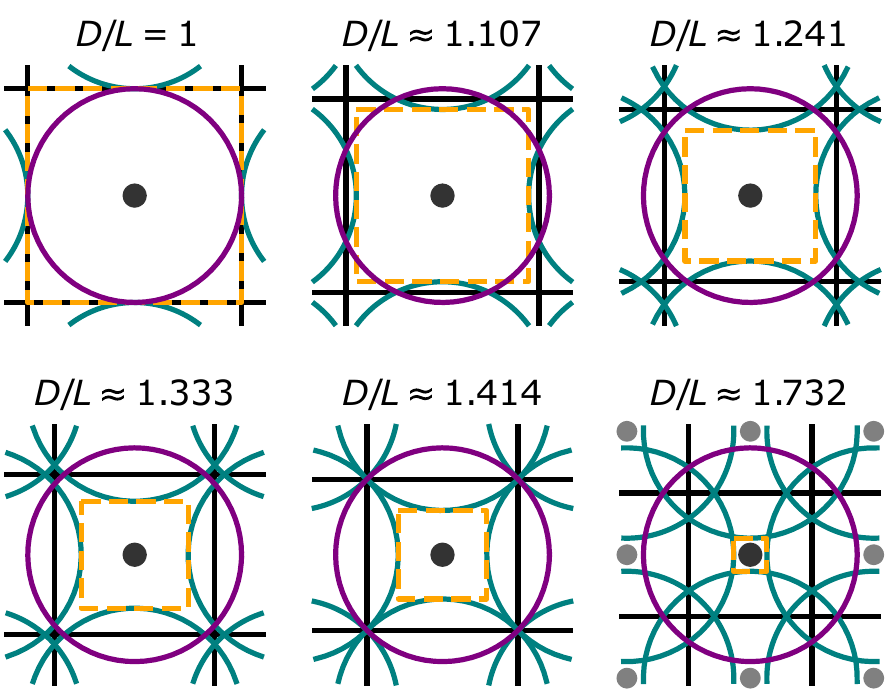}\\
\caption{2D slices through the center of an atom's sphere of influence of diameter, $D$, are shown in relation to periodic simulation boxes with varying edge length, $L$. The atom is represented with a grey dot, and it's periodic images are represented with a light grey dots. The purple circle represents the atoms interaction range and the teal circles/arcs represent the interaction range of the atom's periodic images. The orange dashed line represents the lower bound approximation for the region around the atom that contains no periodic image or ghost atoms.}
    \label{fig:soi_and_box_size}
\end{figure}

Structures are generated via random insertion until a target volume fraction is reached defined by the sum of spherical atomic volumes defined by their respective radii. 
The target atomic volume fraction and elemental composition are randomly chosen from a supplied range of values and used to compute the number of atoms of each element that will be inserted.
Targeting low volume fractions yields low density disordered structures, which may represent gasses or isolated atomic clusters or molecules. 
Targeting high volume fractions results in amorphous solids or glasses. 
To maintain reasonable minimum separation distances while allowing small atomic overlaps, a second set of smaller hard cut-off radii are used for insertion tests.
Atoms are inserted in random elemental order to avoid creating structures where small atoms are densely packed in the voids between larger atoms as is seen in Apollonian sphere packing. 
Although a closed-packed structure represents the ideal limit at a volume fraction of $\pi/({3\sqrt{2}}) \approx 0.74\ldots$ for uniform atom sizes, we find that achieving a volume fraction greater than 0.60 is impractical for a simple random insertion algorithm and 0.50 is generally a reliable upper limit.
When choosing the cell size of a random structure relative to the sphere of interaction (SOI) as defined by the MLIP radial cutoff distance, there is a spectrum of choices. 
On one end, a cell which inscribes the SOI, \textit{i.e.}, a cube with edge lengths $L$ equal to the SOI diameter $D$, permits a truly disordered local environment free of periodic image/ghost atoms and represents amorphous structures (\autoref{fig:soi_and_box_size}). 
%
%
On the opposite side of the spectrum, using a cell with edge lengths of half the SOI diameter means \textit{all} neighboring atoms will have at least one periodic image/ghost within the SOI. 
These structures represent a more periodic structure model, but this may not be as desirable since the aforementioned polymorph-derived structures should comprise the expected crystalline phases. 

When considering the computational cost of creating training data, the former case where the cell edge length is equal to the SOI diameter has 8X the volume of the latter case where the cell edge length is half the SOI diameter.
Moreover since the shortest-range interactions dominate in atomic systems, and regions near the cube vertices may be  outside the SOI, a smaller, less costly, simulation cell can reasonably be used without loss of fidelity.
Further analysis of these tradeoffs is provided in \Autoref{sec:metrics}.
We use the equilibrium pair distances found in \autoref{fig:beak_plot} to compute the elemental radii,  which corresponds to 0.613 \AA\ for oxygen and 1.123 \AA\ for zirconium.
For a the minimum separation distance, a hard cutoff of 90\% of elemental radii was used.
The random cells were a 50/50 mix of 13 \AA and 8 \AA boxes. 
We restricted the atomic volume fractions from 0.05 to 0.50 and compositions to be less that 70 at.\% oxygen.  
\autoref{fig:structure_examples}c is an example of a random structure for Zr/O.

\subsection{Dynamics}
For the three dynamics types, molecular dynamics, contour exploration, and the dimer method, DFT calculations were used to create 100 trajectories starting with polymorph-derived structures and another 100 were using random structures. 
For each structure-dynamics pairing, 
50 trajectories are used for training while the other 50 are used for testing. 
Each of the trajectories was limited to 200 hundred iterations or the queue wall clock, whichever came first. 
Since structures within the same trajectory are similar, using training and test sets from different trajectories ensured a more robust analysis. 
Likewise, only every $n$-th image was included in the datasets to reduce redundancy. 
For each dynamics method, the training and test sets have $\approx 2300$ images each.

\subsubsection{Molecular Dynamics}
\textit{Ab initio} molecular dynamics were performed within VASP using the,  Nos\'e-Hoover thermostat ($NVT$) ensemble and a 1.0\,fs time step. 
Initial and final temperatures were chosen randomly between 200\,K and 2000\,K. 
Every fourth image in each trajectory was included in the data. 
%

\subsubsection{Contour Exploration}

Contour Exploration (CE) was performed using ASE invoking VASP as an energy and force calculator since VASP does not have a contour exploration module.\footnote{
A technical consequence of this configuration is that the VASP program must be started and stopped for each structural iteration limiting performance by about a factor of two. 
Hence, every second image was included rather than every 4th image as with MD.} 
%
Much like with temperature in MD, it is preferable to sample a range of energies using contour exploration. 
A caveat of using this potentiostatic scheme, however, is that there is no way to know the local energy minimum \textit{a priori}. Therefore, we perform a short structural relaxation and then perform contour exploration with an always increasing target potential energy.  
More specifically, each initial structure was relaxed for ten iterations using the RMM-DIIS optimizer in VASP with a step scale of 0.3 for added stability. 
Contour exploration was then performed with incremental ramping from initial and to final potential energy targets, both chosen randomly from 0.001-1.0 eV/atom above the potential energy of the semi-relaxed structure.
%
%
The step size was 0.8 \AA\ and the parallel drift fraction was 0.2 for contour exploration. 

\subsubsection{Dimer-Method Search}

We use the dimer method implemented in the VTST \cite{dimer_vtst} extension to VASP. 
Unlike in molecular dynamics, which has inherently physically meaningful sampling, or contour exploration, where the potentiostat acts to prevent sampling high energy structures outside the range of usefulness, the dimer method searches can easily find high energy saddle points if they are associated with the local minimum eigenmode. 
To partially remedy this, we perform the same relaxation procedure as with CE trajectories so that the dimer-method searches are more likely to find the softer modes near local minima. 
Once relaxed, a random initial displacement of norm 1.2 \AA\ is applied to the atoms and is used as the initial dimer-mode direction. The dimer image spacing was 0.01 \AA\ with a convergence criteria of 0.01 meV/\AA. The GL-BFGS\cite{dimer_vtst} dimer-optimization method in VTST with a curvature damping factor of 4.0 and a maximum step of 0.3 \AA. Every fourth image from the dimer trajectories was added to the datasets. 
%
\section{Results and Discussion}
%
%

%
%
\subsection{Spatial Descriptor Diversity}

\begin{figure}
\centering
\includegraphics[width=0.45\textwidth]{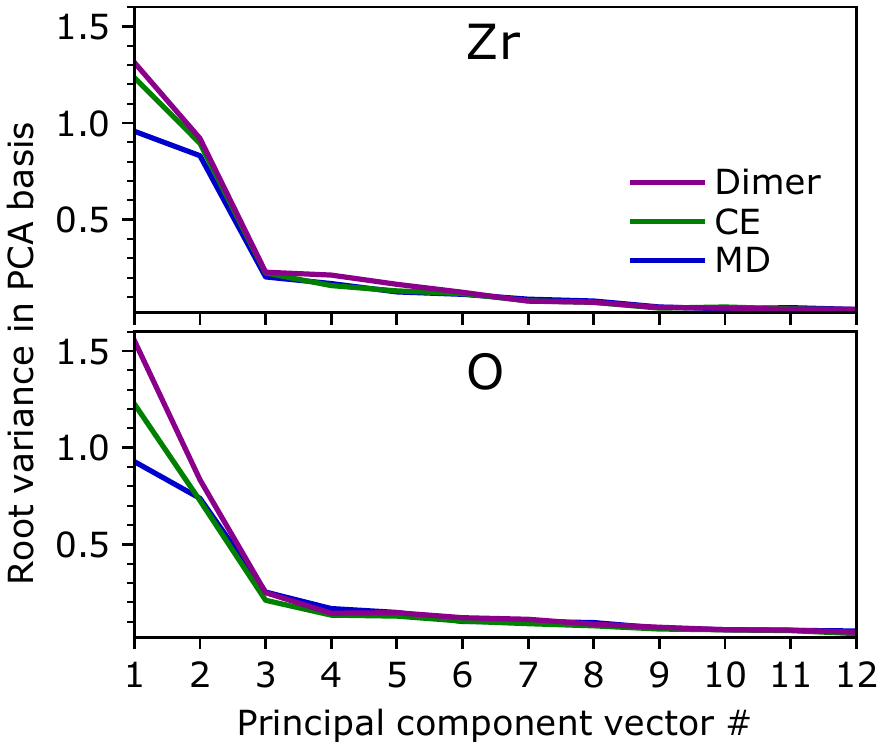}
\caption{Variances of the spatial descriptors in the principal component basis plotted in decreasing order for the first 12 principle components for (upper) Zr and (lower) O.
%
%
}
    \label{fig:pca}
\end{figure}

\begin{figure}
\centering
\includegraphics[width=0.45\textwidth]{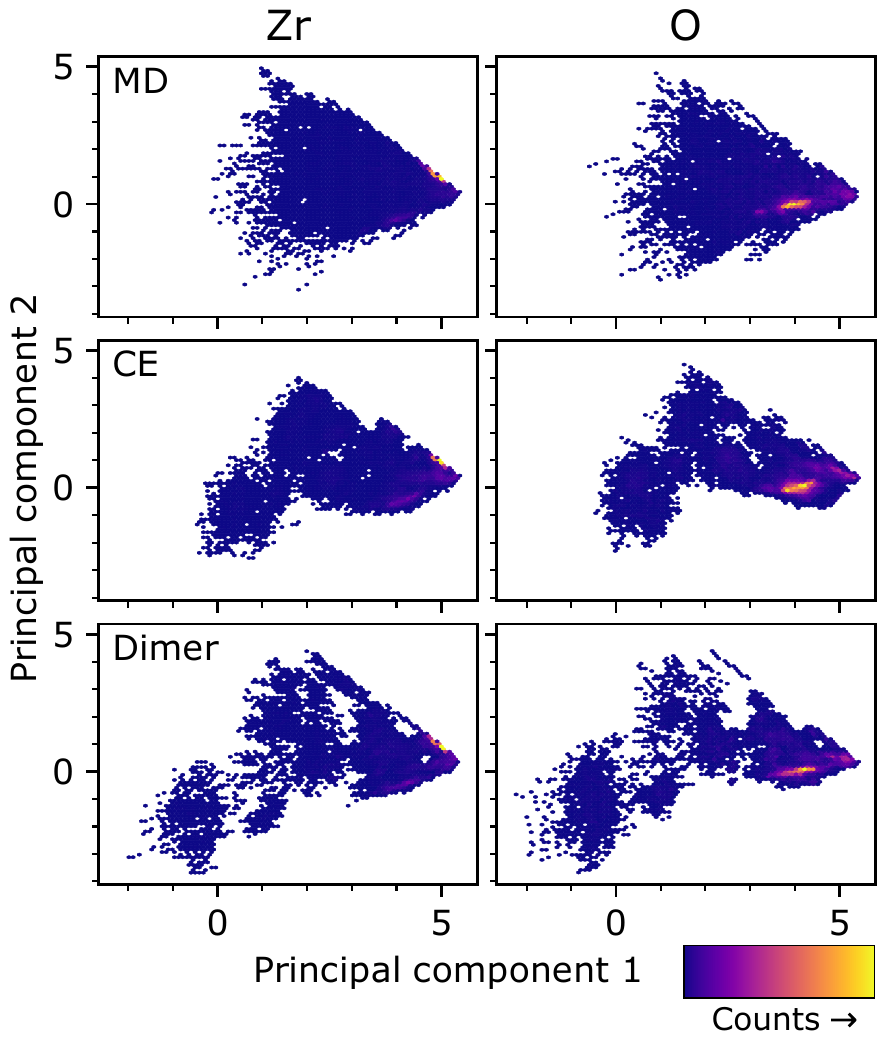}
    \caption{The distributions of the first and second principal components for the MD, CE, and dimer training data sets for  (left column) Zr and (right column) O.
    %
    %
    }
    \label{fig:heatmap}
\end{figure}

To asses the diversity of structures sampled, principal component analysis (PCA) was performed on the combined scaled descriptors across the training sets. 
The descriptors from the different dynamics datasets were projected into the unified PCA basis for a direct comparison. 
The validity of using the same PCA basis across the data sets was confirmed by comparing the $L^{2}$-norms of root-variance of descriptors in the unified PCA basis and the $L^{2}$-norms of the root-eigenvalues of covariance matrices of the individual datasets. The relative difference of the measures does not exceeded $2\times10^{-5}$ in any of the data sets.
%
%
From this analysis, we find the first two principal components are able to  significantly distinguish the diversity of the contour exploration (CE), dimer method (DM), and molecular dynamics (MD) data sets as shown in \autoref{fig:pca}, where the 12 largest of the 136 principle components are shown. 
Furthermore, when comparing the distributions of the first principal components (\autoref{fig:heatmap}), CE and DM exhibit  distributions with similar overall shapes. They exhibit, however,  more widely spread tails and different arrangements of the sample density peaks.  
The overall similarities likely stem from the possible ranges of descriptor being constrained by the spatial geometry of the Behler-Parrinello descriptors.

\begin{figure}
\centering
    \includegraphics[width=0.3\textwidth]{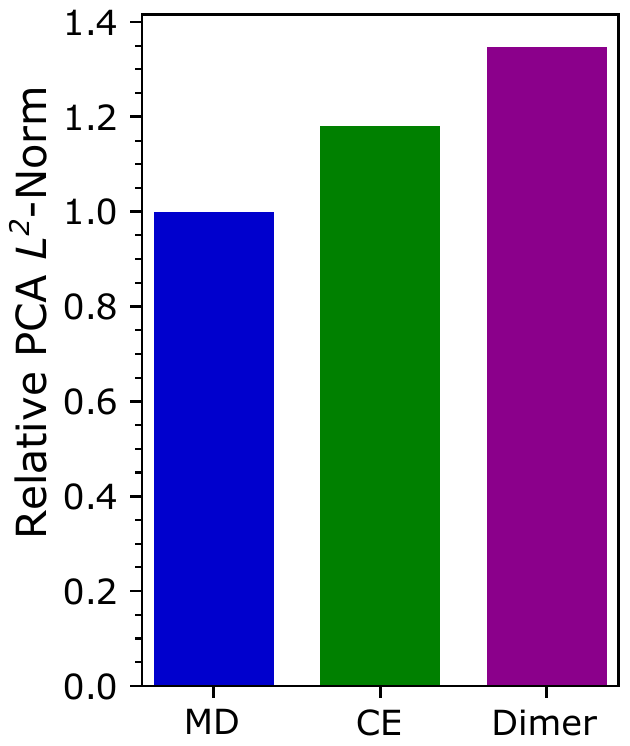}
    \caption{The descriptor principal component root-variances for both zirconium and oxygen are combined in an $L^{2}$-norm to compare data set diversity.  The $L^{2}$-norms are normalized to the MD data set.}
    \label{fig:descriptor_bar}
\end{figure}
To assess the overall descriptor diversity, the $L^{2}$-norm of oxygen and zirconium root-variances in the PCA basis combined in a single vector is used and is show in \autoref{fig:descriptor_bar}, where the values for CE and the dimer method are normalized to the values from MD. 
This analysis yields the length of a box diagonal such that the edges are aligned to the principal component axes of the descriptor distribution and the edges have a length of principal component root-variance. 
By this metric, contour exploration and the dimer method yield an 18\% and 35\% improvements in descriptor diversity, respectively.  

\begin{figure}
\centering
\includegraphics[width=0.45\textwidth]{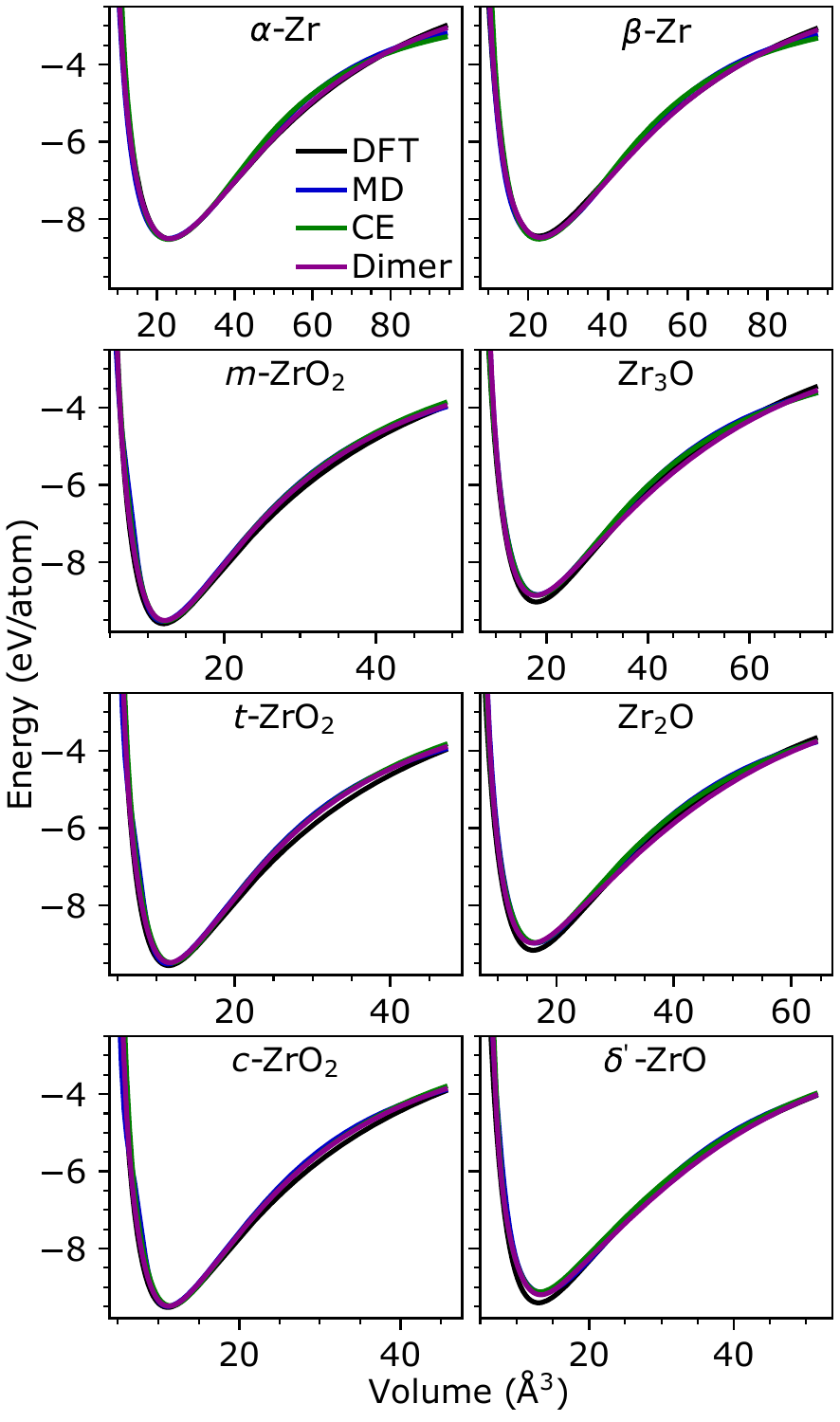}
\caption{Energy-volume curves for Zr metal and {ZrO$_x$} phases, both in the training set, and three suboxide phases not in the training set.}
\label{fig:energy_volume}
\end{figure}

\subsection{Equations-of-state Benchmarks}

Comparing energy calculated from DFT and predicted by a MLIP for a series of isotropic volume dilations (energy-volume curves) for phases in the polymorph-derived structure set (or other known phases) constitutes useful and simple to implement benchmark.
Firstly, it smoothly varies the amount of activation of spatial descriptors from highly activated descriptors to only lowly activated long-range descriptors, essentially sampling the extremes of the Behler-Parrinello Neural Network (BPNN). 
Secondly, the energy-volume data points near the equilibrium volume ($\Omega_0$) can be used to fit an equation of state.
Here we use the simple Murnaghan form, which also provides the bulk modulus.  

\begingroup\squeezetable
\centering
\begin{table}
\caption{Bulk properties calculated from equation of state fitting for the Zr-metal and {ZrO$_x$} phases in the training set and three suboxide phases that were not in the training set.}
\begin{ruledtabular}
\begin{tabular}{lllll}
                 &      \multicolumn{4}{c}{Equilibrium Volume (\r{A}$^{3}$/atom)}    \\ 
\cline{2-5}
 Phase           &            DFT &             MD &             CE &          Dimer \\
 \hline
 $\alpha$-Zr     &          23.46 & 22.87 (-2.5\%) & 23.06 (-1.7\%) & 23.33 (-0.6\%) \\
 $\beta$-Zr      &          22.65 & 22.73 (+0.4\%) & 22.71 (+0.3\%) & 23.09 (+2.0\%) \\
 m-ZrO$_2$       &          12.05 & 11.96 (-0.7\%) & 12.28 (+1.9\%) & 12.23 (+1.4\%) \\
 t-ZrO$_2$       &          11.56 & 11.46 (-0.8\%) & 11.91 (+3.0\%) & 11.69 (+1.1\%) \\
 c-ZrO$_2$       &          11.18 & 11.17 (-0.1\%) & 11.53 (+3.1\%) & 11.29 (+0.9\%) \\
 Zr$_3$O         &          18.05 & 18.29 (+1.4\%) & 18.35 (+1.7\%) & 17.91 (-0.8\%) \\
 Zr$_2$O         &          16.03 & 16.43 (+2.5\%) & 16.45 (+2.6\%) & 16.03 (-0.0\%) \\
 ${\delta}'$-ZrO &          13.04 & 13.61 (+4.3\%) & 13.26 (+1.6\%) & 13.37 (+2.5\%) \\
\hline
\hline\\[-1em]
                 &                \multicolumn{4}{c}{Bulk modulus (GPa)}            \\
 \cline{2-5}
 Phase           &            DFT &             MD &             CE &          Dimer \\
 \hline
 $\alpha$-Zr     &           89.8 &   78.1 (-13\%) &    84.4 (-6\%) &    90.5 (+1\%) \\
 $\beta$-Zr      &           88.7 &   77.0 (-13\%) &    85.8 (-3\%) &    89.2 (+1\%) \\
 m-ZrO$_2$       &          219.1 &   237.5 (+8\%) &   225.0 (+3\%) &   222.5 (+2\%) \\
 t-ZrO$_2$       &          224.5 &  258.1 (+15\%) &   241.2 (+7\%) &   240.2 (+7\%) \\
 c-ZrO$_2$       &          233.2 &  279.5 (+20\%) &  262.6 (+13\%) &  258.4 (+11\%) \\
 Zr$_3$O         &          138.2 &   133.3 (-4\%) &   133.8 (-3\%) &   127.9 (-7\%) \\
 Zr$_2$O         &          159.1 &   157.6 (-1\%) &   160.6 (+1\%) &   152.8 (-4\%) \\
 ${\delta}'$-ZrO &          199.8 &  161.9 (-19\%) &   197.8 (-1\%) &   198.7 (-1\%) \\
\hline
\hline\\[-1em]
                 &          \multicolumn{4}{c}{Relative Energy (meV/atom)}          \\
\cline{2-5}
 Phase           &            DFT &             MD &             CE &          Dimer \\
 \hline
 $\alpha$-Zr     &            0.0 &            0.0 &            0.0 &            0.0 \\
 $\beta$-Zr      &           66.7 &   24.2 (-64\%) &    8.5 (-87\%) &   34.8 (-48\%) \\
 m-ZrO$_2$       &            0.0 &            0.0 &            0.0 &            0.0 \\
 t-ZrO$_2$       &           36.1 &   14.2 (-61\%) &   41.7 (+16\%) &   28.1 (-22\%) \\
 c-ZrO$_2$       &           70.1 &   32.6 (-54\%) &   45.2 (-35\%) &   18.0 (-74\%) \\
 Zr$_3$O         &            0.0 &            0.0 &            0.0 &            0.0 \\
 Zr$_2$O         &         -132.2 &  -138.1 (+4\%) & -107.5 (-19\%) & -113.7 (-14\%) \\
 ${\delta}'$-ZrO &         -374.3 & -299.6 (-20\%) & -266.5 (-29\%) & -333.1 (-11\%) \\
\end{tabular}
\end{ruledtabular}
\label{tab:bulkphase}
\end{table}
\endgroup

All three MLIPs reproduce the shape of the energy-volume curves for the phases in the polymorph-derived generating set, $\alpha$-Zr, $\beta$-Zr, m-ZrO$_2$, t-ZrO$_2$, and c-ZrO$_2$, particularly near the equilibrium volume (\autoref{fig:energy_volume}). 
For the phases not in the polymorph-derived generating set,  ${\delta}^\prime$-ZrO,  Zr$_2$O,  and Zr$_3$O, all three MLIPs reproduce the shape of the EOS well, but the potential energy minima are over-predicted,  \emph{i.e.}, the potential well is not as deep as the DFT result.
Despite the misalignment of the potential energy and thus relative formation energy, the predicted equilibrium volumes and bulk moduli of the phases not in the polymorph-derived generating set were not found to be significantly worse than those predicated for phases within the test set as shown more clearly in \autoref{tab:bulkphase}. 
When comparing the performance of the three MLIPs, their performance in predicting the equilibrium volume is similar. %
The largest percentage errors are 4.3\%, 3.1\%, and 2.5\% for the MD-, CE-, and DM-trained MLIPs, respectively. 
However, for bulk moduli prediction, the DM-trained lead followed by the CE-trained and then MD-trained MLIPs.
In terms of polymorph formation energy, the DM-trained MLIP was best for $\alpha$-Zr, $\beta$-Zr whereas the CE-trained MLIP performed best for \textit{m}-ZrO$_2$, \textit{t}-ZrO$_2$, and \textit{c}-ZrO$_2$. 
\subsection{Vacuum Pair Interactions}
Much like the energy-volume curves, calculating potential energy versus radial separation distance for isolated pairs of atoms is simple, computationally inexpensive, and automatable benchmark. 
Although more relevant to gas-phase phenomena, it is an interesting benchmark since only two-body descriptors are active.  

\begin{figure}
\centering
\includegraphics[width=0.4\textwidth]{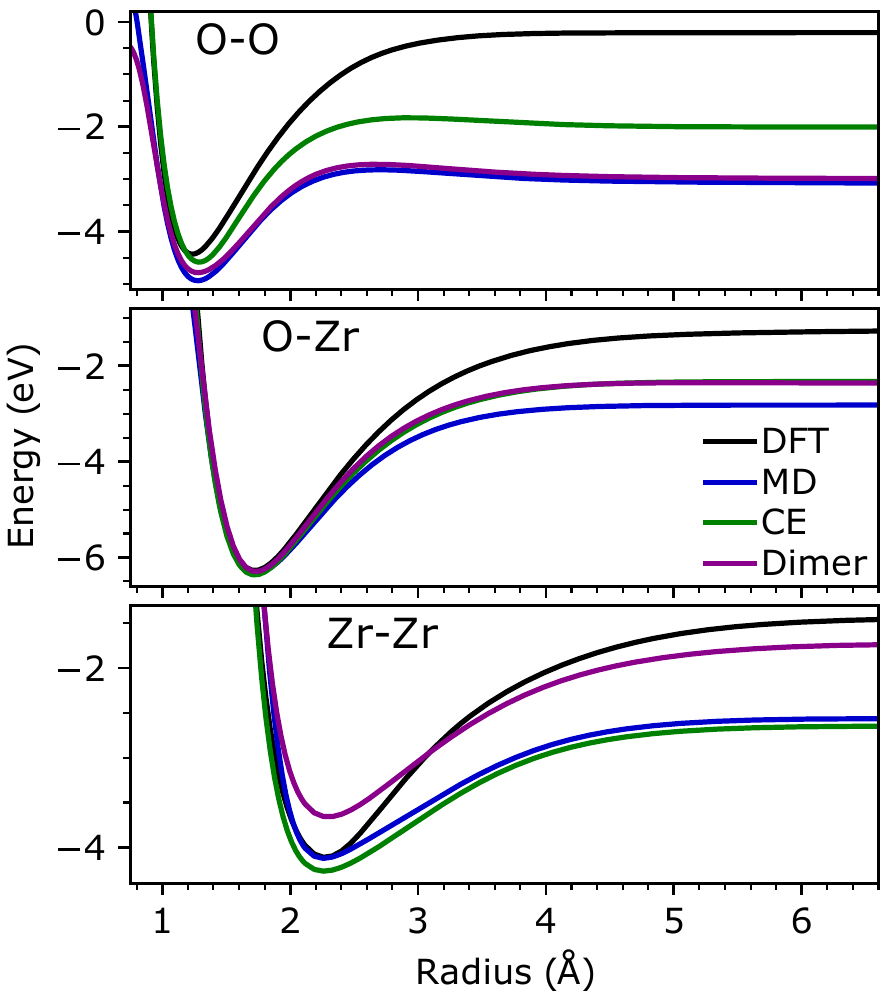}
\caption{Interaction pair potentials for isolated atomic pairs in a vacuum.}
    \label{fig:energy_radius} 
\end{figure}

For the three elemental pairings in the Zr/O system, all three MLIPs produce physically reasonable potential energy curves, and in particular, all three  reproduce the potential well shape of the Zr-O pair accurately (\autoref{fig:energy_radius}). 
Since there are very few isolated atoms in the training set, only being found in the random structures, the binding energies of the pairs are poorly referenced to the vacuum level as shown in \autoref{tab:pair_strength}.
The O-O pair interaction is only qualitatively correct for all three MLIPs, which is due to a lack of explicitly creating oxygen molecules in the training sets.\footnote{Since the training data is non-spin polarized DFT calculations, the reference oxygen molecule is in the singlet state.}
We hypothesize that the better performance of CE for diatomic oxygen is due to the mild tendency of CE to localize motion in isolated rotating oxygen molecules in the lower density random structures.
For the Zr-Zr interactions, all three MLIPs produce similarly-shaped, over-smoothed potential energy wells. 
The Zr-Zr potential energy curve produced by dimer-trained MLIP seemingly sacrifices the equilibrium energy for improved vacuum level alignment. 
This may be related to lack of energy or temperature regulation in the dimer method, since the method simply follows the lowest eigenmode up or down towards a saddle point without energy bounds. 
%
%
Which, given the stiffness of the Zr-O bond, could mean the Zr-Zr bonds are more likely to be broken.
%

%

\begin{table}
\centering
\caption{Atomic binding energies (in eV) for isolated atomic pairs in a vacuum.}
\begin{ruledtabular}
\begin{tabular}{lllll}
Pair  & DFT           & MD              & CE              & DM           \\
\hline
O-O   & 4.23  & 1.86 (-56\%) & 2.58 (-39\%) & 1.80 (-57\%) \\
O-Zr  & 5.01  & 3.50 (-30\%) & 4.04 (-19\%) & 3.93 (-22\%) \\
Zr-Zr & 2.64  & 1.56 (-41\%) & 1.62 (-39\%) & 1.92 (-27\%) \\
\end{tabular}
\end{ruledtabular}
\label{tab:pair_strength}
\end{table}

\subsection{Diffusion Pathways}

\begin{figure}
    \centering
        \includegraphics[width=0.49\textwidth]{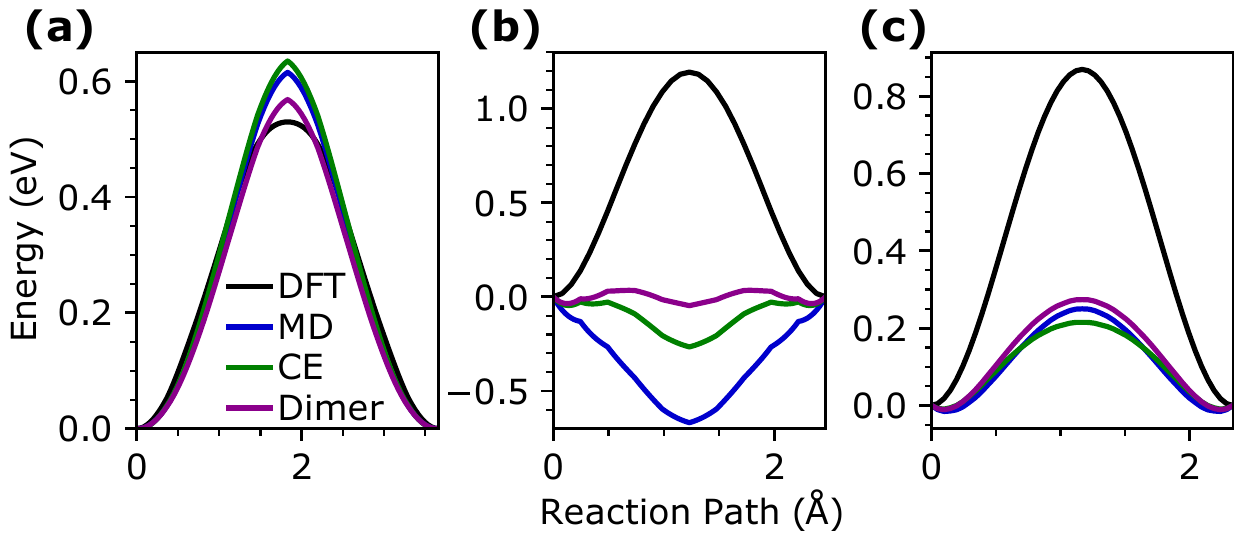}
    \caption{(a) Minimum energy pathway for vacancy diffusion in a $2\times2\times2$ $\alpha$-Zr (HCP) supercell between adjacent basal planes. Oxygen vacancy diffusion MEP in \textit{c}-ZrO$_{2}$ along the [100] direction for (b) a $2\times2\times2$  conventional supercell  with all but the four nearest neighbor Zr atoms fixed and (c) the conventional cell used by Ref.\ \onlinecite{Wang2018a}. 
    }
    \label{fig:diffusion_barriers} 
\end{figure}

Diffusion pathways are challenging to sample for MD since the area of most interest, the saddle point of the transition state, is dynamically unstable and akin to an inverted pendulum. 
Nonetheless, the importance of reproducing the diffusion barrier makes them an useful benchmark. 
Using the nudged elastic band method, the minimum energy pathway (MEP) was calculated at the DFT level.
Rather than compare four different MEPs, we evaluate the MLIPs along the DFT MEP since, an ideally perfect MLIP would have the same MEP as DFT. 
Yet in practice, an imperfect MLIP will have a MEP differing from the DFT result. 
Therefore, the DFT MEP is not guaranteed to pass through the MLIP MEP saddle points, which leads to apparent derivative discontinuities rather than the smooth peak seen in the DFT results (\autoref{fig:diffusion_barriers}).

Across the three pathways shown in \autoref{fig:diffusion_barriers}, the dimer-trained MLIP performs the best. 
This performance is expected because the purpose of the dimer-method search is to find saddle points, but disappointing because none of the MLIPs perform well in the diffusion pathway benchmark and only qualitative accuracy is achieved in HCP-Zr.
In the MLIP literature, it is common for MLIPs to perform well in reproducing reaction pathways for diffusion and stacking fault migration in metallic systems, but less so for insulating or molecular systems.\cite{FeMarzari, NiCuLiMoSiGeOng, TungstenNordlund, TaThompson, TitaniumWen2021,  CarbonCsanyi}
Some authors have taken to more advance ML techniques to address the problems of reaction pathways in molecular systems such as the recent work by Sun \textit{et al.}\cite{SunKozinskyReactions}
These short-comings will need to be addressed before MLIPs can be reliably used to treat transition state phenomena, which motivates exploring other methods such as active and adversarial learning in the future.

\subsection{Phonons}
\begin{figure}
\centering
\includegraphics[width=0.48\textwidth]{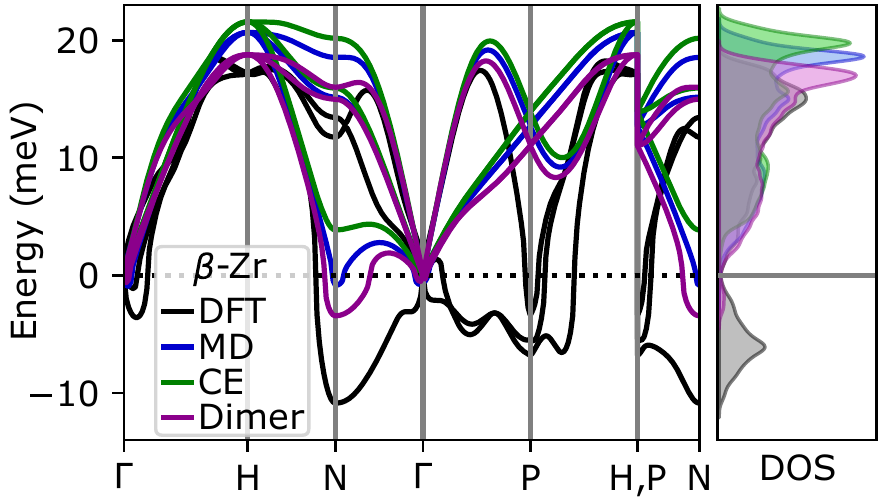}
\caption{Phonon dispersions and vibrational density of states (DOS) for $\beta$-Zr (BCC) calculated using DFT and the three MLIPs. Modes with negative energies are imaginary (unstable) modes.}
\label{fig:phonons}
\end{figure}

Phonon dispersions are sensitive to small changes and/or errors in forces and are key ingredients in thermal transport and harmonic transistion state theory.
Practically, this makes phonon dispersions challenging and important benchmarks for MLIPs.
Here, we compare the phonon dispersions for $\beta$-Zr computed using the super-cell method with a $6\times6\times6$ supercell. 
$\beta$-Zr (BCC) is dynamically unstable at low temperatures\cite{Zr_stability} which is represented by the imaginary modes in \autoref{fig:phonons}. 
A clear ranking of fidelity is observed where the dimer trained MLIP performs best followed by the MD and CE trained MLIPS. 
We attribute the better performance of the MLIP trained using the dimer method due to it being a minimum-eigenmode-following method, which attempts to find saddle points with a pair of system images with a small finite displacement between them. 
This approach is similar to the finite difference scheme used to construct interatomic force constant matrices that are then used to obtain the dynamical matrix and phonon dispersions.
The CE-trained MLIPs predicts no imaginary modes, which is likely due to the range of potential energies targeted, \emph{i.e.}, between 1 meV/atom and 1 eV/atom, with the majority of energies far larger than typical vibrational modes.

\subsection{Cross-Evaluation Statistics}

\begin{figure}
\centering
\includegraphics[width=0.49\textwidth]{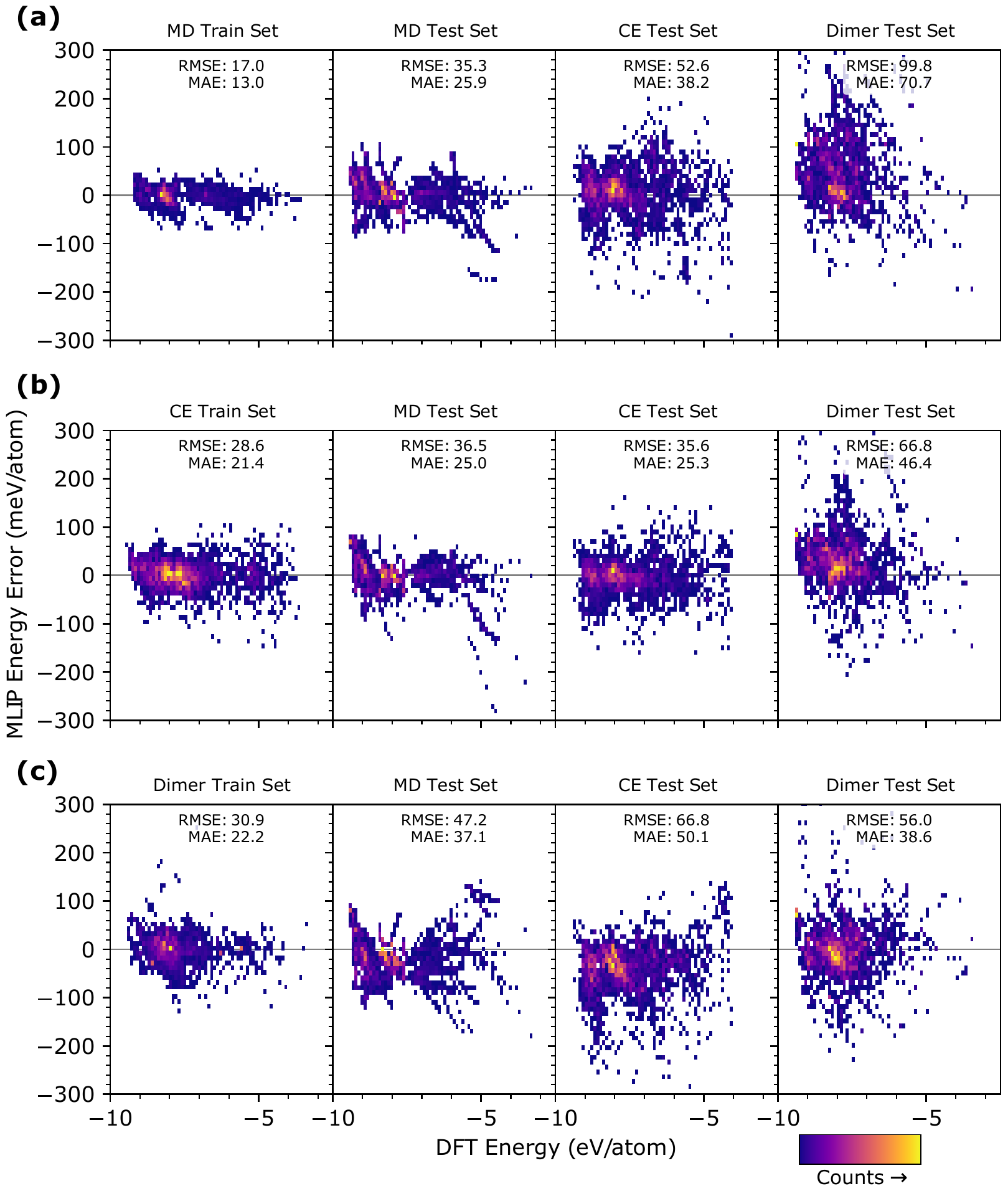}
    \caption{Heatmaps of energy error versus DFT energy for the (a) MD-trained, (b) CE-trained, and (c) DM-trained MLIPs for their own training sets and the three test sets.}
    \label{fig:energy_stats}
\end{figure}

\begin{figure}
\centering
\includegraphics[width=0.49\textwidth]{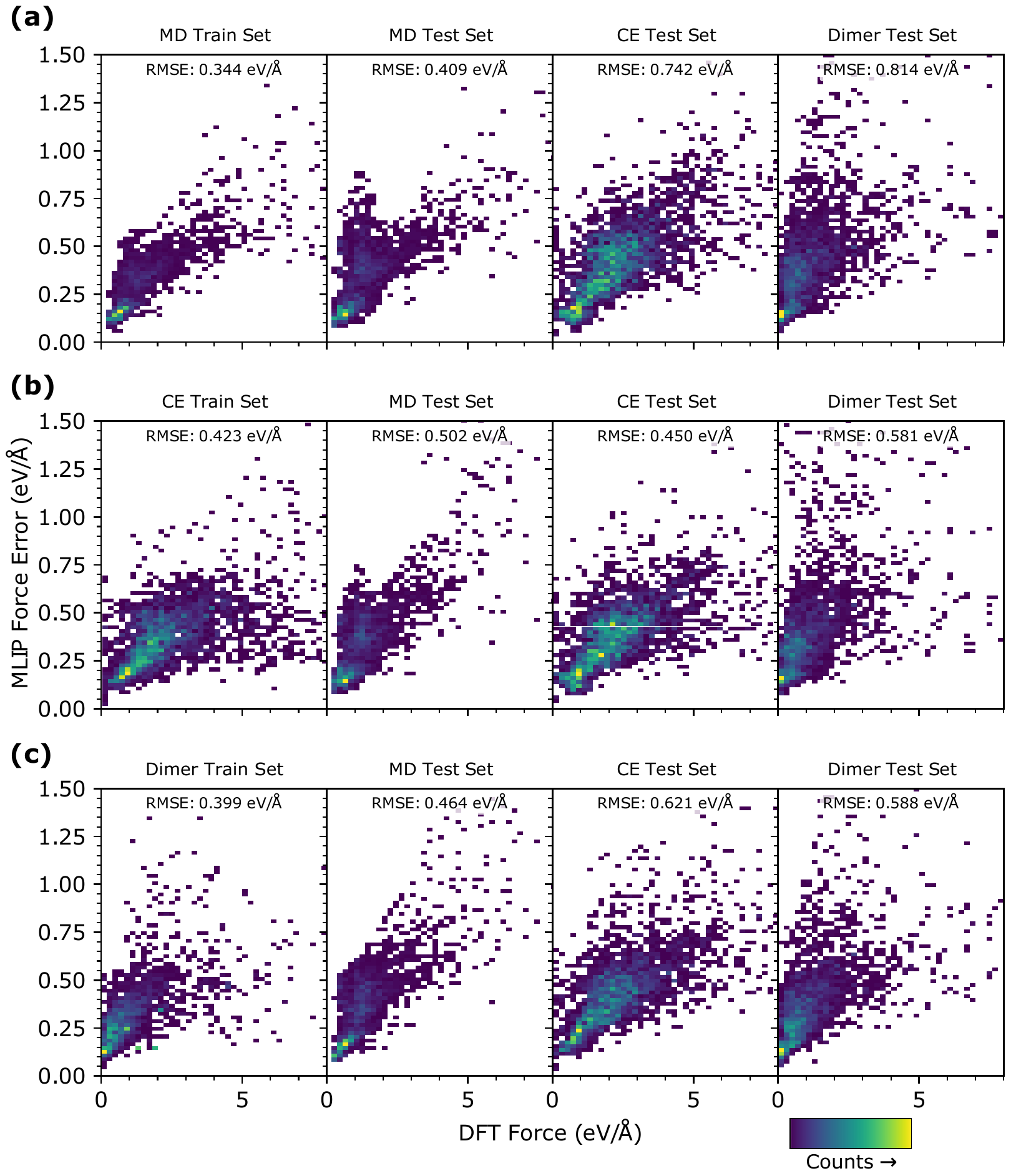}
\caption{Heatmaps of RMS force error versus norm of DFT forces for the (a) MD-trained, (b) CE-trained, and (c) DM-trained MLIPs for their own training sets and the three test sets.}
    \label{fig:force_stats}
\end{figure}

\begin{figure}
\centering
\includegraphics[width=0.49\textwidth]{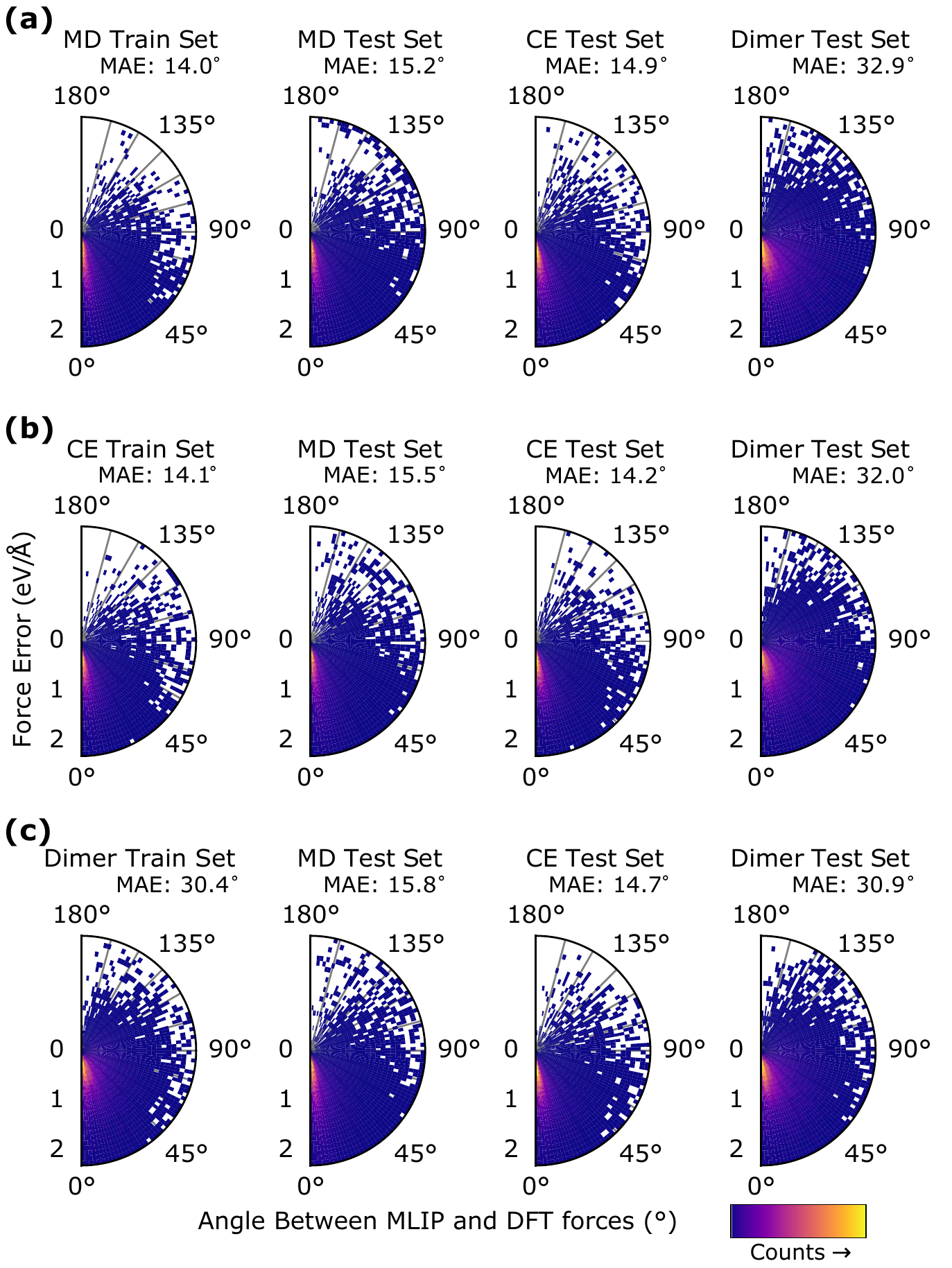}
\caption{Polar heatmaps of the angle between MLIP-predicted and DFT forces versus the magnitude of the atomic force for the (a) MD-trained, (b) CE-trained, and (c) DM-trained MLIPs for their own training sets and the three test sets. The plots are cropped to force magnitude to 2.3\,eV/\AA\ to better show the peak of the distribution.}
    \label{fig:force_stats_polar}
\end{figure}

To further explore the accuracy and generalizable nature of the three MLIPs, we perform cross-validation and cross-test set evaluation. 
If all the data sets were equally diverse in their samplings of atomic environments, then all of the MLIPs would be expected to perform equally well when evaluated on the test set generated by a different type of dynamics, \emph{i.e.}, MD, CE, or dimer, as when they are evaluated on their own test set.
We find this is not the case as see in the different performances in energy accuracy (\autoref{fig:energy_stats}), force-vector accuracy (\autoref{fig:force_stats}), and force-direction accuracy  (\autoref{fig:force_stats_polar}). 
Because forces involve a magnitude and direction, we examine both contributions.
To assess force direction accuracy in a scale-free way, we define the force-angle error, $\theta _\mathbf{F}$, as the angle between the DFT and MLIP atomic force vectors:\[
\theta _\mathbf{F} = \cos^{-1} \left (   \frac{\mathbf{F}_{MLIP} \cdot \mathbf{F}_{DFT} }{\left | \mathbf{F}_{MLIP}  \right | \left | \mathbf{F}_{DFT}  \right |} \right )\,,
\]
where $\mathbf{F}_{DFT}$ and $\mathbf{F}_{MLIP}$ are  the force vectors from DFT and those predicted by the MLIP, respectively.
In the broader machine-learning community, this is  essentially a cosine similarity metric.\cite{CosineSimilarity} 
As an example, a 10$^{\circ}$ force error on a 1.0 eV/\AA\ force vector has approximately the same RMS error as a 180$^{\circ}$ force error on a 0.05 eV/\AA\ force. 
For many properties such as phonons dispersions, the accuracy of the small forces near equilibrium are more important than far from equilibrium. As such, future loss functions could incorporate the metric so that training includes force direction in a way that is more disconnected from the overall force magnitudes.
Starting with energy and force accuracy, the best overall performance for any test set was achieved with the MLIP trained on the same dynamics type, unsurprisingly. 
The best performance for the MD- and CE-trained MLIPs was on the matching test sets, while the DM-trained MLIP performed better on the MD test set than on the dimer test set. 
Similarly, the CE-trained MLIP performed similarly well on the MD test set as the CE test set. 
The force angle errors in \autoref{fig:force_stats_polar} are dominated by the dynamics used to generate the data set rather than the evaluated MLIP's original training set dynamics type;  the mean-force-angle errors are 14$^{\circ}$-16$^{\circ}$ for the CE and MD data sets while the mean-force-angle errors are much higher at 30$^{\circ}$-33$^{\circ}$ for dimer data sets regardless of the MLIP used.  
We suspect this is a result of the dimer rotation steps giving spatially similar structures but with force vector variations.
When coupled with the higher descriptor diversity of the DM data set, it suggests that increasing the MLIP model complexity might permit better treatment of the DM data regardless of the dynamics used to generate the training set.
Overall, the dimer data presents striking differences from the CE and MD data sets for the purposes of training MLIPs.

The much poorer  energy and force accuracy performance of the MD-trained MLIP on the CE- and DM-trained test sets is almost certainly due to the poorer data set diversity, which we previously showed through the principal component analysis. 
The lower data set diversity of MD is further highlighted by the fact that individual trajectories can be discerned in energy error heatmaps for the MD test and train sets in \autoref{fig:energy_stats}.  
This is apparent despite including only every fourth image from the MD trajectories. 
While a seemingly good solution would be to increase the MD timestep or temperature to increase the distance between iterations, these choices have to be checked for stability with regards to integrating the equations of motion. 
Furthermore, for a mixed element system, \emph{e.g.}, hydrogen carbons and transition metals, motion of the lightest and fastest elements will dominate the exploration of the PES\cite{Respa} since the average velocity varies as $ \langle v \rangle \propto \sqrt {T/m}$ in the Maxwell–Boltzmann distribution.
In contrast, the CE and DM do not need timestep tuning, are independent of element mass, and have sampling schemes which only depend on the shape of the PES. 


\begin{figure}
\centering
\includegraphics[width=0.49\textwidth]{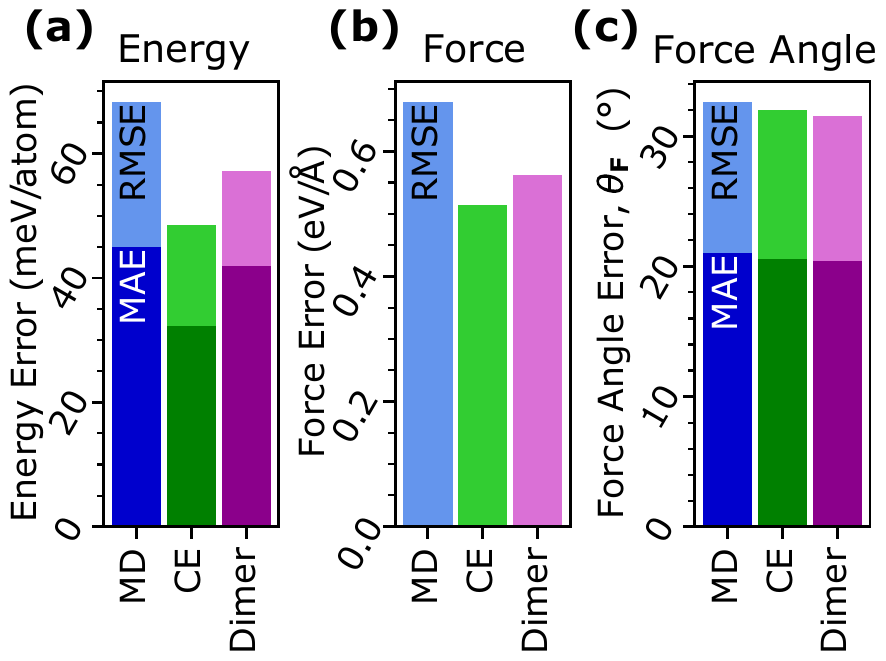}
\caption{Cross-test set evaluation summary for (a) energy error, (b) force vectors, and (c) angle error of force vectors, where light areas are RMSE dark are MAE.}
\label{fig:stat_summary}
\end{figure}

%
Since each type of atomic motion has biases in the types of structures that are generated, the MLIPs trained on one type of motion generally perform well on the test set for that type of motion than MLIPs trained on other motion types.
To assess the overall generalizability, we compare the statistical metrics for each MLIP when evaluating all three tests sets simultaneously. 
In this manner, the MLIPs are evaluated for the exact same structures.  
In these combined metrics, the CE-trained MLIP has the best general performance in combined energy and forces tests and the dimer-trained MLIP has the best performance in the combined force angle error test (\autoref{fig:stat_summary}).  
From these overall metrics, the cross test-set evaluations, and the various benchmarks, it's clear that MD is not as efficient as CE or the dimer method when used as a means of sampling structures for training MLIPs. 
While CE statistically outperforms the dimer method searches for training MLIPs, the performance in benchmarks and unusual nature of the structures generated suggests the a mix dimer searches and CE would be better yet.

\section{Conclusions}
Molecular dynamics, contour exploration, and dimer method searches were evaluated and cross-evaluated for their ability to create diverse training data sets for machine-learned interatomic potentials. 
Despite its common use in generating training data, we found MD was outperformed in every statistical evaluation and almost every benchmark by contour exploration, and dimer method searches. 
In our overall evaluation, we find the CE is the most efficient for generating training data for MLIPs. 
However, dimer-method searches should be also be considered for future training set generation because ($i$) it yielded the most descriptor-diverse training data, ($ii$) the dimer-trained MLIP was the least erroneous in diffusion MEPs, and ($iii$) the dimer-trained MLIP more closely reproduced the phonon dispersions of $\beta$-Zr.
Thus we suggest future works use a combination of dimer searches and CE, where CE uses shallower potential energy ramping to better sample low energy structures. 
In building towards a fully automatable MLIP training scheme, where automatically created initial structures are evolved using dynamics methods to sample more configuration space, robust structure generations schemes are still in need of improvement. 

\begin{acknowledgments}
The authors thank Nicholas Wagner for code testing and enlightening discussions, Vishu Gupta for his valuable feedback and discussions, and the ASE project\cite{ase-paper} for greatly simplifying the implementation of contour exploration and the data management for this work.
This work relates to Department of Navy Award N00014-20-1-2368 issued by the Office of Naval Research. The United States Government has a royalty-free license throughout the world in all copyrightable material contained herein.
Computational efforts were supported by the National Energy Research Scientific Computing Center (NERSC), a U.S. DOE Office of Science User Facility located at Lawrence Berkeley National Laboratory, operated under Contract No. DE-AC02-05CH11231, and the Center for Nanoscale Materials (Carbon) Cluster, an Office of Science user facility supported by the U.S. Department of Energy, Office of Science, Office of Basic Energy Sciences, under Contract No. DE-AC02-06CH11357. 
\end{acknowledgments}

\section*{Code Availability}
The structure-creation code is currently available in a beta state at \url{https://github.com/MTD-group/amlt}.

\appendix
\renewcommand{\thefigure}{A\arabic{figure}}
\renewcommand{\thetable}{A\arabic{table}}
\setcounter{figure}{0}
\setcounter{table}{0}

\section{Gaussian Descriptors\label{sec:descriptors}}
The Behler-Parrinello style Gaussian descriptors $G^{(2)}_i$ for atom $i$ take the form
\begin{align}
G_{i}^{(2)}=\sum_{j\ne i}{e^{-\eta R_{ij}^2/R_c^2}f_c\left(R_{ij}\right)},
\label{eq:G2}
\end{align}
where the sum runs over atoms $j$ of radial distance $R_{ij} < R_c$ of atom $i$. 
The $G^{(4)}_i$ descriptors take the form
\begin{align}
G_{i}^{(4)}=2^{1-\zeta}\!\!\sum\limits_{\substack{j,\,k\ne i \\
(j\ne k)}}
{F_c^{(4)}\left(1+\gamma \cos \theta_{ijk}\right)^\zeta
e^{-\eta\left(R_{ij}^2+R_{ik}^2+R_{jk}^2\right)/R_c^2}}
\label{eq:G4}
\end{align}
where $F_c^{(4)}=f_c\left(R_{ij}\right)f_c\left(R_{ik}\right)f_c\left(R_{jk}\right)$, and the sum is over atoms $j$ and $k$ within the distance $R_c$ of atom $i$.
The $G^{(5)}_i$ descriptors take the form
\begin{align}
G_{i}^{(5)}=2^{1-\zeta}\!\!\sum\limits_{\substack{j,\,k\ne i \\
(j\ne k)}}{F_c^{(5)}\left(1+\gamma \cos \theta_{ijk}\right)^\zeta
e^{-\eta\left(R_{ij}^2+R_{ik}^2\right)/R_c^2}}
\label{eq:G5}
\end{align}
where $F_c^{(5)}=f_c\left(R_{ij}\right) f_c\left(R_{ik}\right)$, and the sum is over atoms $j$ and $k$ within the distance $R_c$ of atom $i$.

\begingroup
\squeezetable
\begin{table*}
    \centering
    \caption{Various metrics for the uniqueness and cost of random structures taken from geometric considerations. The simulation cell sizes are given by edge length, $L$, the diameter of the sphere of influence (SOI) is $D$, and the SOI is the interaction volume defined by the interaction cutoff radius, $r_{cut}$ such that $D=2r_{cut}$. An inscribed sphere has $D=L$.} 
\begin{ruledtabular}
\begin{tabular}{p{1.5in}p{1in}p{1in}p{1in}p{1.in}p{1in}}
\multirow{3}{0.8in}{\centering Geometric Description}
 & \multirow{3}{0.8in}{\centering\normalsize$\frac{\textrm{Diameter}}{\textrm{Length}}$}  & \multicolumn{4}{c}{Metrics (Descriptor Ratios)}  \\
\cline{3-6}
&
&\multirow{2}{1in}{(a) $\frac{\textrm{SOI Volume}}{\textrm{Cell Volume}}$}  
&\multirow{2}{1in}{(b) $\frac{\textrm{Unique Radius}}{\textrm{Cutoff Radius}}$} 
&\multirow{2}{1in}{(c) $\frac{\textrm{Unique Volume}}{\textrm{SOI Volume}}$}
& \multirow{2}{1in}{(d) Relative Cell Volume}  \\ 
& & & & & \\
  \hline
Inscribed Sphere              &                                                                1     &                                                            $\pi/6 \approx 0.524$  &                                                            1  &                                                                  $6/\pi \approx 1.910$     &                                                                        1    \\
Unique volume = SOI volume   &  $\frac{2}{\left(\frac{\pi}{6}\right )^{\frac{1}{3}}+1}\approx 1.107$ &    $288\pi/\left(\pi^{\frac{1}{3}} 6^{\frac{2}{3}} +6 \right )^{3} $ &  $\frac{ 6^{\frac{2}{3}} \pi^{\frac{1}{3}}}{6} $ &                                                                                   1       &  $\left(\pi^{\frac{1}{3}} 6^{\frac{2}{3}} +6 \right )^{3}/1728 $ \\
Cell volume = SOI volume              &              $\left(\frac{6}{\pi}\right )^\frac{1}{3}\approx 1.2407$ &                                                                                  1 & $\frac{ 6^{\frac{2}{3}} \pi^{\frac{1}{3}}}{3}-1$ &  $\frac{\left( 2\pi^{\frac{1}{3}} - 6^{\frac{1}{3}} \right)^{3}}{\pi} $ &                                                        $\pi/6 $  \\
Inner half of $r_{cut}$ is unique &                                                $4/3 \approx 1.333$   &                         $\left(\frac{4}{3}\right )^{3} \frac{\pi}{6}$ &                                           $1/2 $ &                                                       $\frac{3}{4\pi} $      &                                $\left(\frac{3}{4}\right )^{3} $ \\
SOI touches cell edges                &                                                   $\sqrt{2} = 1.414$ &                                              $\frac{\pi \sqrt{2}}{3} $ &                                          $\sqrt{2}-1$ &                                              $\frac{ 30\sqrt{2} - 42}{\pi} $ &                            $\frac{1}{\left(\sqrt{2}\right )^{3}} $ \\
SOI circumscribes cell                &                                                   $\sqrt{3} = 1.732$ &                                               $\frac{\pi\sqrt{3}}{2} $ &                         $\frac{2}{\sqrt{3}} - 1 $ &                                          $ \frac{ 52/\sqrt{3} - 30}{\pi} $ &                          $\frac{1}{\left(\sqrt{3}\right )^{3}} $  \\
\end{tabular}
\end{ruledtabular}
\label{tab:random_size}
\end{table*}
\endgroup

\section{Random Structure Uniqueness and Cost Metrics \label{sec:metrics}}

To help balance computational cost and the sampling of unique local environments with few periodic images atoms, we formulated and tabulated six useful metrics (\autoref{tab:random_size}). 
Metric (a) is the ratio of the SOI volume to cell volume. This approximately  represents the ratio of the useful part to costly part since any ML model is only informed by atoms within the SOI volume. 
Metric (b) is $r_{unique}/r_{cut}$, whereby we start with the maximum radius where all neighboring atoms are guaranteed to be unique (without periodic images), or the unique radius, $r_{unique}$ and normalize it to the cut-off radius $r_{cut}$.
In \autoref{fig:soi_and_box_size}, the unique radius is the distance from the center to the nearest point on the adjacent periodic image SOI, \emph{i.e.},  $r_{unique} = L - r_{cut}$.

\begin{figure}
\centering
\includegraphics[width=0.42\textwidth]{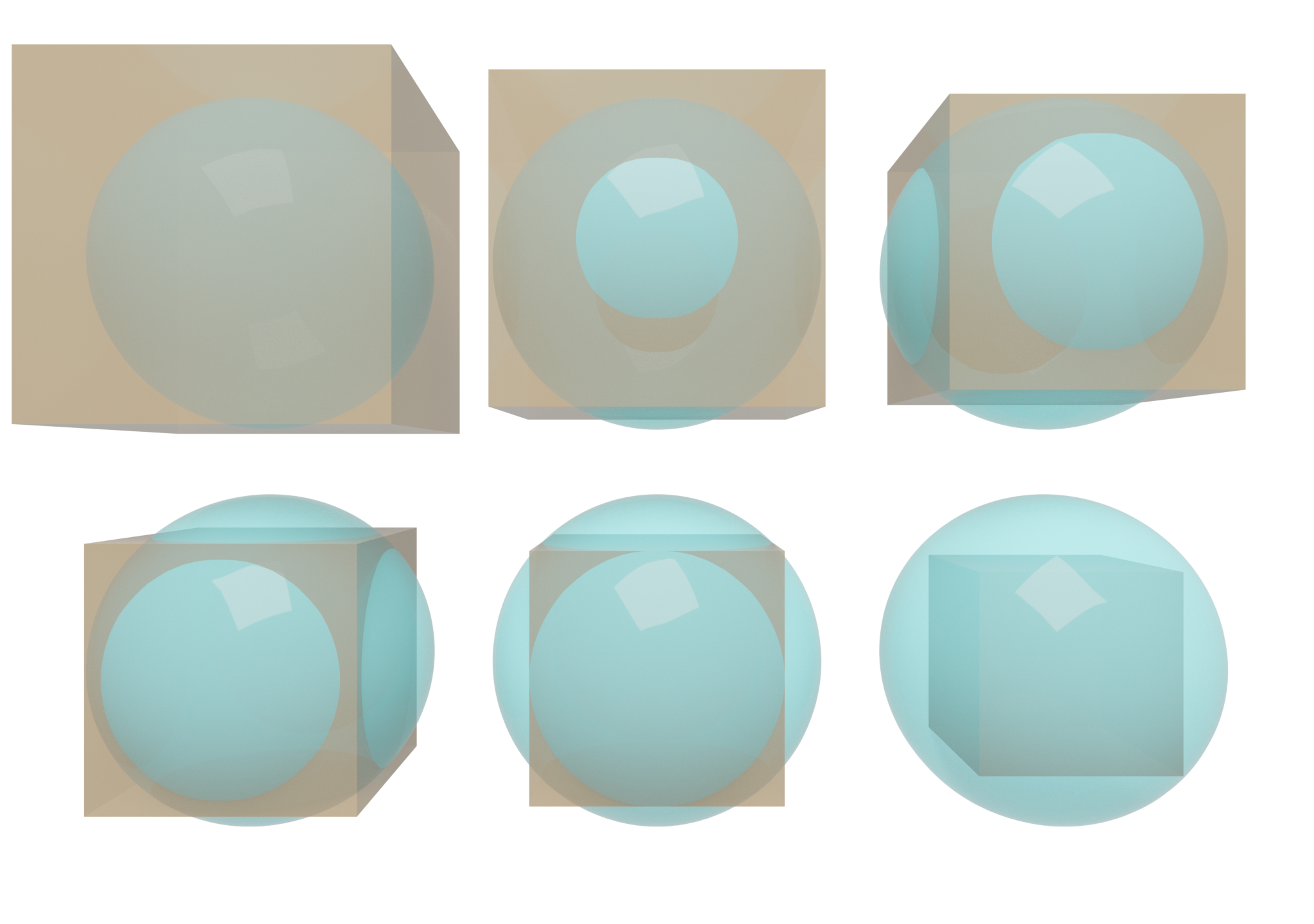}
\caption{The six tabulated sphere of influence (SOI) diameter-to-simulation-cell-size ratios appearing in \autoref{fig:soi_and_box_size} rendered with fixed SOI diameter.}
    \label{fig:rendered_soi_and_box_size}
\end{figure}

Next, we estimate a lower bound of the volume where all neighbors are guaranteed to be unique with a cube that extends from the center of the cell and just touches the adjacent periodic image SOI. 
The unique volume cube has edge length $2 \times r_{unique}$. 
Its perimeter is represented by the dashed line in \autoref{fig:soi_and_box_size}, 
which we further render in three-dimensions in \autoref{fig:rendered_soi_and_box_size}.
Metric (c) is then the unique volume to SOI volume ratio, $= \left ( 2 \: r_{unique} \right )^{3}/({\frac{4}{3} \pi r_{cut}^{3}})$, and is an approximate lower bound on the fraction of neighbors that are guaranteed to be unique. 
Metric (d) is the cell volume relative to the cell volume of the most costly scenario, \emph{i.e.}, a SOI inscribed in the cell or $D/L=1$, and is calculated as $(D/L)^{-3}$.  
This metric helps assess the computational cost.
Based on the values in \autoref{tab:random_size}, and in the absence of benchmarks, it would seem that most reasonable $D/L$ ratios are in the range of 1.1-1.4 but $D/L=1$.  
%
%

The choice of interaction sphere diameter relative to random box size reduces to the balance of representing more truly random atomic environments and computational cost.
Here there are additional tradeoffs: For gas phases, sampling weaker longer range interactions (such as van der Waals) is desirable, so larger, low density cells are required. 
Conversely, short-range screening in solids, particularly metals, may alleviate some of the need for truly random environments. 
Thus in it maybe be possible find simple heuristic which minimizes cost by using sparsely filled large cells and densely filled small cells.

%

\bibliography{references}

\end{document}